%% file: paper_irs_ce_fesl_arxiv.tex
\documentclass[conference]{IEEEtran}
\IEEEoverridecommandlockouts
\usepackage{cite}
\usepackage{amsmath,amssymb,amsfonts}
\usepackage{algorithmic}
\usepackage{graphicx}
\usepackage{textcomp}
\usepackage{xcolor}
\usepackage{pgfplots}
\pgfplotsset{compat=1.15}
\usepgfplotslibrary{statistics}
\usetikzlibrary{patterns}
\usetikzlibrary{decorations.pathmorphing,calc,shapes,shapes.geometric}
\usepgfplotslibrary{groupplots,dateplot}
\usetikzlibrary{patterns,shapes.arrows}
\usetikzlibrary{backgrounds,positioning,fit,matrix,fillbetween}
\usepackage{bm}
\usepackage{tabularx}
\usepackage{subcaption}
\usepackage{balance}

\input{befe}

\newacronym{gmm}{GMM}{Gaussian mixture model}
\newacronym{pdf}{PDF}{probability density function}
\newacronym{cme}{CME}{conditional mean estimator}
\newacronym{siso}{SISO}{single-input single-output}
\newacronym{pdp}{PDP}{power delay profile}
\newacronym{dps}{DS}{Doppler spectrum}
\newacronym{fft}{FFT}{fast Fourier transform}
\newacronym{ofdm}{OFDM}{orthogonal frequency-division multiplexing}
\newacronym{ml}{ML}{machine learning}
\newacronym{mt}{MT}{mobile terminal}
\newacronym{bs}{BS}{base station}
\newacronym{ul}{UL}{uplink}
\newacronym{dl}{DL}{downlink}
\newacronym{fdd}{FDD}{frequency division duplex}
\newacronym{tdd}{TDD}{time division duplex}
\newacronym{ce}{CE}{channel estimation}
\newacronym{los}{LOS}{line-of-sight}
\newacronym{nlos}{NLOS}{non-line-of-sight}
\newacronym{o2i}{O2I}{outdoor-to-indoor}
\newacronym{uma}{UMa}{urban macrocell}
\newacronym{ls}{LS}{least squares}
\newacronym{ris}{RIS}{reconfigurable intelligent surface}
\newacronym{simo}{SIMO}{single-input multiple-output}
\newacronym{ula}{ULA}{uniform linear array}
\newacronym{ura}{URA}{uniform rectangular array}
\newacronym{awgn}{AWGN}{additive white Gaussian noise}

\def\BibTeX{{\rm B\kern-.05em{\sc i\kern-.025em b}\kern-.08em
		T\kern-.1667em\lower.7ex\hbox{E}\kern-.125emX}}

\pgfplotsset{tick label style={font=\small},label style={font=\small},legend style={font=\scriptsize}}

\definecolor{myblack}{RGB}{70,70,70}
\definecolor{myblue}{RGB}{65,105,225}
\definecolor{mygreen}{RGB}{0,139,139}
\definecolor{myorange}{RGB}{255,150,0}
\definecolor{myred}{RGB}{255,69,0}
\definecolor{mylila}{RGB}{153,50,204}

\newcommand{\quadriga}{QuaDRiGa}

\newcommand{\marksize}{1.6pt}

\tikzset{genie/.style={mark options={solid},color=black, line width=\lineWidth}}
\tikzset{global/.style={mark options={solid},color=TUMBeamerGreen, line width=\lineWidth, mark=triangle, mark size=\marksize, dashed}}
\tikzset{gmm/.style={mark options={solid},color=TUMBeamerBlue, line width=\lineWidth, mark=square, mark size=\marksize}}
\tikzset{gmm_diag/.style={mark options={solid},color=TUMBeamerLightBlue, line width=\lineWidth, mark=square, mark size=\marksize, dotted}}
\tikzset{gmm_kron/.style={mark options={solid},color=black, line width=\lineWidth, mark=square, mark size=\marksize, dotted}}
\tikzset{2x1gmm/.style={mark options={solid},color=mylila, line width=\lineWidth, mark=diamond, mark size=\marksize}}
\tikzset{2x1gmm_mid/.style={mark options={solid},color=TUMBeamerGreen, line width=\lineWidth, mark=triangle, mark size=\marksize, dashed}}
\tikzset{2x1gmm_mid_circ/.style={mark options={solid},color=TUMBeamerRed, line width=\lineWidth, mark=x, mark size=\marksize, dashdotted}}
\tikzset{old/.style={mark options={solid},color=TUMBeamerYellow, line width=\lineWidth, mark size=\marksize, dashdotted}}
\tikzset{em_gm_gamp/.style={mark options={solid},color=TUMBeamerRed, line width=\lineWidth, mark=triangle, mark size=\marksize, dashdotted}}
\tikzset{double_gmm_genie/.style={mark options={solid},color=gray, line width=\lineWidth, mark=triangle, mark size=\marksize, dashdotted}}
\tikzset{pdp2x1/.style={mark options={solid},color=gray, line width=\lineWidth, mark=o, mark size=\marksize, solid}}
\tikzset{pdp_kron/.style={mark options={solid},color=black, line width=\lineWidth, mark=o, mark size=\marksize, dotted}}
\tikzset{gmm_toep/.style={mark options={solid},color=TUMBeamerGreen, line width=\lineWidth, mark=x, mark size=\marksize, dashed}}

\begin{document}
	
	\title{Channel Estimation with Reduced Phase Allocations in RIS-Aided Systems
		\thanks{
			The authors acknowledge the financial support by the Federal Ministry of Education and Research of Germany in the programme of ``Souver\"an. Digital. Vernetzt.''. Joint project 6G-life, project identification number: 16KISK002.
		}
}
	\author{
		\centerline{Benedikt Fesl, Andreas Faika, Nurettin Turan, Michael Joham, and Wolfgang Utschick}\\
		\IEEEauthorblockA{School of Computation, Information and Technology, Technical University of Munich, Germany\\
			Email: \{benedikt.fesl, andreas.faika, nurettin.turan, joham, utschick\}@tum.de\\
		}
	}
	
	\maketitle
	
	\begin{abstract}
		We consider channel estimation in systems equipped with a \ac{ris}. In order to illuminate the additional cascaded channel as compared to systems without a \ac{ris}, commonly an unaffordable amount of pilot sequences has to be transmitted over different phase allocations at the \ac{ris}. 
		However, for a given \ac{bs} cell, there exist immanent structural characteristics of the environment 
		which can be leveraged to reduce the necessary number of phase allocations.
		We verify this observation by a study on \ac{dft}-based phase allocations where we exhaustively search for the best combination of \ac{dft} columns. 
		Since this brute-force search is unaffordable in practice,
		we propose to learn a \ac{nn} for joint phase optimization and channel estimation because of the dependency of the optimal phase allocations on the channel estimator, and vice versa. 
		We verify the effectiveness of the approach by numerical simulations where common choices for the phase allocations and the channel estimator are outperformed.
		By an ablation study, the learned phase allocations are shown to be beneficial in combination with a different state-of-the-art channel estimator as well.
	\end{abstract}
	
	\begin{IEEEkeywords}
		Reconfigurable intelligent surface, channel estimation, phase optimization, convolutional neural network.
	\end{IEEEkeywords}

	\section{Introduction}
	\ac{ris}-aided systems are enabling enhanced communication performance
	and are thus considered to be a key technology in 6G systems \cite{IRS_6G}.
	Having accurate estimates of both the direct and the cascaded channel including the \ac{ris} is crucial.
	Since the \ac{ris} only consists of passive elements, processing the impinging waves is not possible.
	Consequently, no separate channel estimation can be conducted at the \ac{ris}.
	To fully illuminate the cascaded channel, a large number of training sequences has to be transmitted over different phase allocations at the \ac{ris}.
	However, the transmission of these long training sequences drastically diminishes the available time for data transmission and hence decreases the achievable rate. Since the number of training phase allocations scales both in the number of \ac{ris} elements and in the number of pilot sequences for a MIMO system, it is generally considered to be unaffordable to fully illuminate the cascaded channel \cite{9722893}.

	
	A variety of approaches for phase optimization and channel estimation is considered in the literature. An on/off strategy was proposed in
	, e.g. \cite{9130088}, where the direct and the cascaded channels are estimated subsequently, which is known to be suboptimal \cite{9053695}. 
	As shown in \cite{9747624,9053695}, the \ac{dft} matrix is the optimal phase allocation matrix when employing the \ac{ls} estimator for full illumination. 
	Unfortunately, the optimal phase allocations are unknown in general for the \ac{mmse} channel estimator or when having reduced phase allocations.
	Therefore, some prior works have considered phase optimization or reduction for specific instances.
	The work in \cite{9133142} discusses optimization of discrete phase shifts, and \cite{9081935} investigates joint pilot and phase optimization for the \ac{mmse} estimator in the full illumination case. 
	In \cite{9543577}, a projected gradient descent algorithm for optimizing the phase allocation matrix is proposed which only holds for a sparse geometry-modeled channel. 
	An element grouping strategy was proposed in \cite{9039554} whose disadvantage is the loss of degrees of freedom at the \ac{ris}.
	To summarize, none of the existing prior works considers phase reduction and optimization jointly for a generally unknown and arbitrary complex channel distribution which we consider in this paper.
	
	\textit{Contributions:}
	We investigate the optimization of phase shifts at the \ac{ris} for channel estimation inside a specific \ac{bs} cell, especially for a reduced number of phase allocations.
	To analyze the potential of phase optimization with respect to a given radio propagation environment, we perform a study on reduced \ac{dft}-based phases where we exhaustively search for the best combination of \ac{dft} columns as phase matrix, which is shown to be heavily dependent on the considered scenario.
	
	Motivated by this observation, we propose a \ac{nn} which jointly learns the phase matrix and the channel estimator. 
	The first part of the \ac{nn} emulates the observed signal by interpreting the angles of the reduced phase matrix as parameterizable weights. The phase matrix module by design fulfills the unit magnitude constraint enforced by the passive nature of the \ac{ris} elements that is problematic in classical \ac{ris} optimization algorithms. 
	This allows to adjust the reduced phase matrix to the propagation scenario by training.
	 The second part of the \ac{nn} consists of a \ac{cnn} for channel estimation.
	We show in numerical experiments that the proposed approach outperforms \ac{dft}-based and random phase allocations together with state-of-the-art channel estimators. 
	We further perform an ablation study to evaluate the properties of the optimized phase allocations, i.e., the performance with respect to a different channel estimator.


\section{System and Channel Model}\label{sec:system}
We consider a \ac{ris}-aided single-input multiple-output (SIMO)\glsunset{simo} system where we denote the direct channel between a single-antenna \ac{mt} and an $M$-antenna \ac{bs} by $\B h_0\in\C^M$. The channel between the \ac{ris} with $L$ passive elements and the \ac{mt} is denoted by $\B h_1\in\C^L$, whereas the channel between the \ac{ris} and the \ac{bs} is denoted by $\B H_2\in\C^{M\times L}$. The received uplink signal is then given by
	\begin{align}
		\B y^\prime &= \B h_0 + \B H_2\B \Phi \B h_1 + \B n^\prime
		\label{eq:system1}
	\end{align}
where $\B \Phi = \diag(\B v)\in\mathbb{C}^{L\times L}$ comprises the unimodular phase shift coefficients at the \ac{ris} elements and $\B n^\prime\sim\mathcal{N}_\C(\B 0, \sigma^2\eye)$ is \ac{awgn}. 
Due to the passive elements at the \ac{ris}, the amplitudes of the reflected signals are not changed. Hence, $v_\ell= \op e^{\op j \theta_\ell}$
with the angle $\theta_\ell\in[0,2\pi)$ and unit-magnitude entries $|v_\ell| = 1$ for $\ell=1,\dots,L$.
With $\B H = [\B h_0, \B h_1\T \circledast \B H_2 ]\in\C^{M\times L+1}$, where $\circledast$ denotes the Khatri-Rao product,
the system in \eqref{eq:system1} can be written as $\B y^\prime = \B H \B v^\prime + \B n^\prime$ where $\B v^\prime = [1, \B v\T]\T$, see e.g., \cite{9747624}.

Note that $\B h_1$ and $\B H_2$ of the cascaded channel $\B H_2\B \Phi \B h_1$ cannot be estimated explicitly \cite{9747624}. Therefore, $N_v$ different phase allocations are considered, that are collected in $\B V = [\B  v^\prime_1,\dots,\B v^\prime_{N_v}]$, to illuminate the channel. This yields 
\begin{align}\label{eq:system_unvec}
	\B Y &= \B H \B V +  \B N \in\C^{M\times N_v}
\end{align}
as the training sequence where the  $N_v$ different observations are collected as the columns of $\B Y$.
After vectorization, we get
\begin{equation}
	\B y = (\B V\T \otimes \eye) \B h + \B n = \B A \B h + \B n \in\C^{MN_v},
	\label{eq:system_vec}
\end{equation}
with the vectorized expressions $\B h = \vect(\B H)$, $\B y=\vect(\B Y)$, $\B n=\vect(\B N)$, and the observation matrix $\B A = \B V\T \otimes \eye$, where $\otimes$ denotes the Kronecker product.
We define the \ac{snr} as $\text{SNR} = 1/\sigma^2$ where we normalize the channels to $\op E [\|\B h\|_2^2] = M(L+1)$.

For the construction of a scenario-specific channel dataset, we use the \quadriga channel simulator \cite{QuaDRiGa1}.
We consider an \ac{uma} scenario following the 3GPP 38.901 specification, where the \ac{bs} is placed at a height of 25m and covers a sector of 120°. The \ac{ris} is placed opposite to the \ac{bs} with a distance of 500m at the same height
. Note that, opposite to the \ac{mt} with possibly \ac{nlos} channels, the channel between \ac{ris} and \ac{bs} has \ac{los} condition. We want to highlight that, although the position of the \ac{bs} and \ac{ris} is fixed, the corresponding channel is not constant within the dataset but slightly changes according to the \ac{uma} conditions. 
The generated channels are post-processed to remove the path gain.


\section{Reference Methods}\label{sec:refs}

\subsection{Channel Estimation}\label{sec:refs_est}

We briefly introduce the \ac{gmm} and the \ac{cme} based thereon from \cite{9842343,9747226}.
A \ac{gmm} with $K$ components is a \ac{pdf} of the form
$f_{\B h}^{(K)}(\B h) = \sum_{k=1}^K p(k) \mathcal{N}_{\C}(\B h; \B \mu_k, \B C_k)$
consisting of a weighted sum of $ K $ Gaussian \acp{pdf}.
Given data samples, an \ac{em} algorithm can be used to fit a $ K $-components \ac{gmm}~\cite[Sec. 9.2]{bookBi06}.
In \cite{9842343,9747226}, a \ac{cme} is formulated based on a \ac{gmm}, which is proven to asymptotically converge to the true \ac{cme} when $K$ grows large. The estimator is formulated as a convex combination of \ac{lmmse} terms, given as
\begin{equation}\label{eq:gmm_full}
	\hat{\B h}^{(K)} = \sum_{k=1}^K p(k \mid \B y) ( \B \mu_k + \B C_k\B A\h \B C_{\B y,k}\inv (\B y - \B A\B \mu_k))
\end{equation}
where the responsibilities $p(k \mid \B y)$ are computed by 
\begin{equation}\label{eq:gmm_likelihood}
	p(k \mid \B y) = \frac{p(k) \mathcal{N}_{\C}(\B y; \B A\B\mu_k, \B C_{\B y,k}) }{\sum_{i=1}^K p(i) \mathcal{N}_{\C}(\B y; \B A\B \mu_i, \B C_{\B y,i}) }
\end{equation}
with $\B C_{\B y,k} = \B A \B C_k \B A\h +\sigma^2 \eye$, cf. \eqref{eq:system_vec}. 
In order to reduce the online computational complexity of the estimator, structural features of the covariances can be utilized, cf. \cite{gmm_structured}, which is out of the scope of this paper.

A \ac{lmmse} estimator is evaluated as baseline using a cell-wide sample covariance matrix $\B C = \frac{1}{N}\sum_{n=1}^N \B h_n \B h\h_n$ with $N=19\cdot 10^4$ training samples to compute
\begin{equation}
	\B h_{\text{sample cov.}} = \B C\B A\h (\B A \B C \B A\h + \sigma^2\eye)\inv \B y.
	\label{eq:sample_cov}
\end{equation}

Finally, the \ac{ls} estimator is $\hat{\B h}_{\text{LS}} = \B A^\dagger \B y = (\B V^\dagger \otimes \eye)\B y$,
where $\B V^\dagger$ is the pseudoinverse of $\B V$.

\subsection{Phase Allocations}\label{sec:refs_phase}

A simple choice for the phase allocations is to use random phase shifts for every \ac{mt}. We therefore construct a phase matrix by sampling i.i.d. Gaussian realizations from $\mathcal{N}_\C(0,1)$ per entry and dividing each entry by its absolute value to fulfill the unit magnitude constraint. Note that these phase allocations might be difficult to implement in a practical system because of the very limited processing ability at the \ac{ris}.

Since \ac{dft}-based phases are optimal in the full-illumination case for the \ac{ls} estimator  \cite{9747624,9053695}, we evaluate the use of a \ac{dft} submatrix for reduced phase allocations, i.e., the $m,n$th entry is given as $V^{\text{sub-DFT}}_{m,n} = \exp((m-1)(n-1)\op j2\pi/N_v)$ with $m=1,\dots,L+1$ and $n=1,\dots,N_v$.
Note that the columns of the \ac{dft} submatrix are not orthogonal for $N_v < L+1$.

\section{DFT-Based Phase Allocation Study}\label{sec:dft_study}

\begin{figure}[t]
	\centering
	\begin{tikzpicture}
		\begin{axis}
			[
			ybar=-7pt,
			bar width=7pt,
			width=\plotwidth,
			height=0.5\columnwidth,
			xtick={1,3,5,7,9,11,13,15,17},
			xmin=0, 
			xmax=18,
			xlabel={DFT Column},
			ymin= 1e-2,
			ymax=1e-1,
			ylabel= {Relative Frequency}, 
			grid = both,
			legend columns = 3,
			legend entries={
				Parallel,
				30$^\circ$ Downtilt,
			},
			legend style={at={(0.5,1.0)}, anchor=south},
			]
			\addplot[color=TUMBeamerBlue,line width=1.2pt,pattern=north east lines, pattern color=TUMBeamerBlue,opacity=0.7]
			table[x=col, y=rel, col sep=comma]
			{csvdat/histogram-DFT_scen1_1x8BS_1x1UE_4x4IRS_Nv=8_IRS-parallel_SNR=40dBm.csv};
			
			\addplot[mark options={solid},color=TUMOrange,line width=1.2pt,fill=TUMOrange,opacity=0.7,pattern=crosshatch dots, pattern color=TUMOrange,densely dashed]
			table[x=col, y=rel, col sep=comma]
			{csvdat/histogram-DFT_scen2_1x8BS_1x1UE_4x4IRS_Nv=8_30deg-down_SNR=40dBm.csv};
			
			%
			
		\end{axis}
	\end{tikzpicture}
	\caption{Histogram of the occurrence of the \ac{dft} columns in the exhaustive search approach for different \ac{ris} configurations with $M=8$, $L=16$, and $N_v =8$ at a \ac{snr} of 40dBm.}
	\label{fig:histogram_40dBm}
\end{figure}
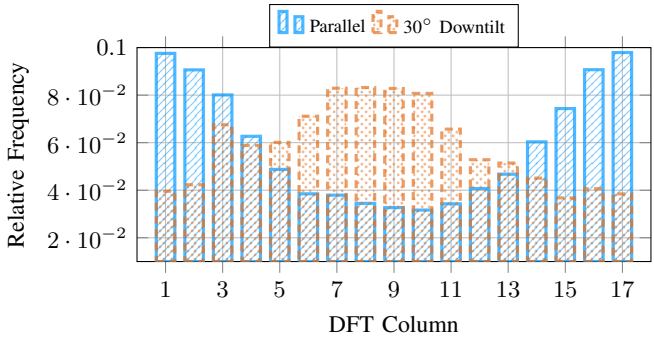

%
%
%

In this section, we investigate the potential of the phase allocation optimization based on a \ac{dft} grid. 
We consider a \ac{bs} with a \ac{ula} consisting of $M=8$ antennas serving single-antenna \acp{mt} supported by a \ac{ris} with $L=4\times 4 = 16$ elements
. Instead of full illumination with $L+1$ phase allocations at the \ac{ris} we set $N_v=8$ in order to simulate a reduced phase allocation situation.
We then exhaustively search for the best combination of eight columns drawn from the full $(L+1)$-dimensional \ac{dft} matrix for $10,000$ uniformly sampled \acp{mt} in the \ac{bs} cell. 
Note that this procedure is infeasible in practical systems since in general $\binom{L+1}{N_v}$
combinations of \ac{dft} columns have to be tested for every \ac{mt} which drastically increases for higher numbers of \ac{ris} patches. 
In the considered case this already yields $\binom{17}{8} = 24,310$ combinations.
For each \ac{mt}, we choose the combination which yields the best channel estimation performance based on the \ac{gmm} estimator introduced in \Cref{sec:refs_est} at an \ac{snr} of 40dBm. The histogram in Fig. \ref{fig:histogram_40dBm} shows how often each \ac{dft} column occurs relatively in the exhaustive search of \ac{dft} column combinations over all \acp{mt} for two different scenarios. 

In the first scenario, the \ac{ris} array is placed in parallel to the \ac{bs} array where it can be observed that especially the first and last \ac{dft} columns occur more frequently in the best-performing combinations. On average, the best combination of \ac{dft} columns for this scenario is $\{1,2,3,4,14,15,16,17\}$. In contrast to that, for a scenario where the \ac{ris} has a downtilt of $30^\circ$, the middle \ac{dft} columns occur primarily in the best combinations and $\{3,5,6,7,8,9,10,11\}$ is the best combination on average.

The conclusions of this study are twofold. First, we have seen that there is great potential for optimizing the phases since for a given setting, some \ac{dft} columns are much more important than others. Second, we showed that the optimization of the phase allocations heavily depends on the considered scenario.
We further evaluate the optimization based on the \ac{dft} grid in the numerical experiments section where we compare this exhaustive brute-force search approach to our proposed optimization procedure.

\section{Learning-Based Joint Phase Optimization and Channel Estimation}\label{sec:joint}

In \Cref{sec:dft_study}, we have seen that the choice of the phase allocation matrix is depending heavily on the underlying system setup, i.e., the configuration of the \ac{ris}, as well as on the propagation environment that induces structural properties which can be exploited for reduced phase allocations. 
However, on the one hand, the optimization procedure from \Cref{sec:dft_study} is infeasible in practice because of the combinatorial search, on the other hand, it is limited to a search on the \ac{dft} grid which may be sub-optimal in general.

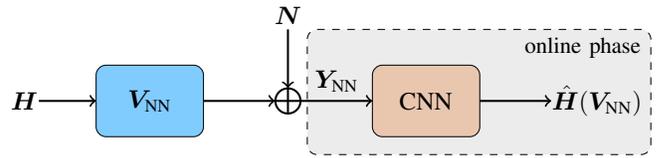
\begin{figure}[t]
	\centering
	\resizebox{1\columnwidth}{!}{
	\begin{tikzpicture}
		\draw[fill=gray,fill opacity=0.15,rounded corners,dashed] (3.95,-0.75) rectangle ++(4.8,1.75);
		\node at (7.8,0.75) {\small{online phase}};
		
		\node at (0,0) {$\B H$};
		\draw [->, thick] (0.2,0) -- (1,0) {};
		\draw[fill=TUMBeamerBlue,fill opacity=0.5,rounded corners] (1,-0.5) rectangle ++(1.5,1);
		\node at (1.75,0) {$\B V_{\text{NN}}$};
		\draw [->, thick] (2.5,0) -- ++(1,0) {};
		\filldraw[black, fill=white,thick] (3.68,0) circle (5pt);
		\draw[thick] (3.5,0) -- ++(0.36,0) {};
		\draw[thick] (3.68,0.18) -- ++(0,-0.36) {};
		\node at (3.68,1.2) {$\B N$};
		\draw [->, thick] (3.68,1) -- ++(0,-0.82) {};
		\draw [->, thick] (3.86,0) -- node[above] {$\B Y_{\text{NN}}$} ++(1,0) {};
		\draw[fill=TUMOrange,fill opacity=0.3,rounded corners] (4.86,-0.5) rectangle ++(1.5,1);
		\node at (5.61,0) {CNN};
		\draw [->, thick] (6.36,0) -- ++(1,0) {};
		\node at (8,0) {$\hat{\B H}(\B V_{\text{NN}})$};
	\end{tikzpicture}
	}	
	\caption{Flowchart of the proposed \ac{nn} architecture for joint phase optimization and channel estimation. 
	In the online phase only the \ac{cnn} is evaluated for a fixed $\B V_{\text{NN}}$.
	}
	\label{fig:flowchart}
\end{figure}

Thus, we propose to utilize machine  learning for joint phase optimization and channel estimation via a specific \ac{nn} architecture in \ac{ris}-aided systems. In essence, we parametrize the phase allocations of the matrix $\B V$. Due to the passiveness of the \ac{ris} which enforces the unit magnitude constraint, we only train with respect to the angles of the phase matrix. In particular, the phase matrix is constructed as 
\begin{equation}\label{eq:V_learning}
	\B V_{\text{NN}} =  \cos(\B \Phi) + \op j \sin(\B \Phi)
\end{equation}
where $\B \Phi\in\R^{L+1\times N_v}$. A similar approach for the optimization of a sensing matrix with a magnitude constraint was employed in \cite{KOLLER2022108553} which serves as a motivation for our considerations.

The training procedure is summarized as follows.
The parametrized phase matrix $\B V_{\text{NN}}$ 
given by \eqref{eq:V_learning} is multiplied with a channel realization from the training dataset. Afterwards, we artificially add \ac{awgn}, yielding an emulated observation $\B Y_{\text{NN}}$ following the model in \eqref{eq:system_unvec}. 
The emulated observation $\B Y_{\text{NN}}$ then serves as the input of a \ac{cnn} which yields a channel estimate $\hat{\B H}(\B V_{\text{NN}})$ at the output. Therefore, the complex-valued input of the \ac{cnn} is split into its real and imaginary part as different convolution channels and each layer employs 2D convolutions.
Since the phase optimization and the training of the \ac{cnn} for channel estimation depend on each other, it is not possible to separately update their parameters. Thus, we jointly optimize the phase matrix $\B V_{\text{NN}}$ and the \ac{cnn} for which we exploit the efficient framework of \acp{nn} with powerful gradient-based optimization techniques. As such, we can interpret the phase matrix $\B V_{\text{NN}}$ as a layer with a specific structure, cf. \eqref{eq:V_learning}, of a larger \ac{nn} that contains the \ac{cnn} as further layers. 
The described architecture is summarized as a flowchart in Fig. \ref{fig:flowchart}. 
We utilize labeled data from the constructed dataset, cf. \Cref{sec:system}, to compute gradients with the \ac{mse} 
\begin{equation}
	\text{MSE} = \op E[ \| \B H - \hat{\B H}(\B V_{\text{NN}}) \|_F^2]
\end{equation}
as cost function. Note that a single forward pass propagates through both \ac{nn} parts and, therefore, all network parameters are updated simultaneously. After training, the optimized phase allocations are given by \eqref{eq:V_learning} and the trained \ac{cnn} is extracted as the channel estimator. 

We initialize the weights of the phase matrix randomly at the beginning of the training and we perform a random hyper-parameter search for the \ac{nn} parameters, i.e., the batch size ($\in[2^5,2^{11}]$), activation functions (ReLU, Tanh, Sigmoid, SiLU, ELU), batch normalization, learning rate ($\in[10^{-5}, 10^{-1}]$), number of kernels ($\in [16, 512]$) and layers ($\in[3,9]$) for $3\times 3$ convolution kernels, where we choose the best setting over 100 random initializations.

The optimized phase matrix $\B V_{\text{NN}}$ after training is further evaluated by an ablation study where it is used in combination with a different channel estimator, i.e., the \ac{gmm} estimator from \Cref{sec:refs_est}, instead of the trained \ac{cnn}. Since the resulting performance is better in comparison to the baseline phase allocations (cf. \Cref{sec:numeric}), although the \ac{gmm} estimator is not jointly trained with the phase matrix, we conclude that the optimized phase matrix $\B V_{\text{NN}}$ is beneficial for the whole \ac{bs} cell and exploits its structural properties.

The online computational complexity of the proposed \ac{cnn} estimator is determined by a single forward pass, which depends on the chosen hyper-parameters, since the optimized phase matrix $\B V_{\text{NN}}$ is fixed after training, cf. Fig. \ref{fig:flowchart}. Note that we also employ the trained phase matrix $\B V_{\text{NN}}$ when using the \ac{gmm} estimator whose online complexity is given in \cite{9842343}, \cite{gmm_structured}.

\section{Numerical Results}\label{sec:numeric}
We present numerical results for the described setting in \Cref{sec:system}. We utilize a dataset consisting of $19\cdot10^4$ data samples for fitting the \ac{gmm} with $K=128$ components and training the \ac{nn}. Each method is evaluated using $10^4$ samples which are not part of the training data. For all plots, we evaluate the scenario with a parallel \ac{ris} with a \ac{ura} opposite to the \ac{bs} with a \ac{ula} since the results are qualitatively the same for both depicted scenarios in \Cref{sec:dft_study}.
The curves labeled ``LS'', ``sample-cov'', or ``GMM'' refer to the baseline estimators from \Cref{sec:refs_est}, whereas ``CNN joint'' refers to the proposed approach from \Cref{sec:joint}. The additional labeling ``DFT'', ``rand'', ``opt'', or ``hist'' refers to the choice of the phase allocation matrix based on the \ac{dft} (sub)matrix or on random allocations, cf. \Cref{sec:refs_phase}, the optimized phase allocations from the \ac{nn}, cf. \Cref{sec:joint}, or the histogram based search from \Cref{sec:dft_study}, respectively. 
	
\subsection{Full Illumination}\label{sec:numeric_full}

\begin{figure}[t]
	\centering
	\begin{tikzpicture}
		\begin{axis}
			[width=1\columnwidth,
			height=0.57\columnwidth,
			xtick=data, 
			xmin=-10, 
			xmax=40,
			xlabel={SNR [dBm]},
			ymode = log, 
			ymin= 1e-3,
			ymax=1e0,
			ylabel= {Normalized MSE}, 
			ylabel shift = 0.0cm,
			grid = both,
			legend columns = 2,
			legend entries={
				\scriptsize LS DFT,
				\scriptsize sample-cov DFT,
				\scriptsize sample-cov rand,
				\scriptsize GMM DFT,
				\scriptsize GMM rand,
				\scriptsize GMM opt,
				\scriptsize CNN joint,
			},
			legend style={at={(0.0,0.0)}, anchor=south west},
			]
			
			\addplot[mark options={solid},color=black,line width=1.2pt,mark=x]
			table[x=SNR, y=LS, col sep=comma]
			{csvdat/2022-10-11_19-42-17_ant_irs=16_ant_bs=8_ant_ue=1_comp=128_sum=0.99_ntrain=190000_ntest=10000_nv=17_np=1_NLOS_dftphase=True_down=False.csv};
			
			
			\addplot[mark options={solid},color=TUMOrange,line width=1.2pt,mark=triangle]
			table[x=SNR, y=sample_cov, col sep=comma]
			{csvdat/2022-10-11_19-42-17_ant_irs=16_ant_bs=8_ant_ue=1_comp=128_sum=0.99_ntrain=190000_ntest=10000_nv=17_np=1_NLOS_dftphase=True_down=False.csv};
			
			\addplot[mark options={solid},color=TUMOrange,line width=1.2pt,mark=triangle,dashed]
			table[x=SNR, y=sample_cov, col sep=comma]
			{csvdat/2022-10-13_05-27-19_ant_irs=16_ant_bs=8_ant_ue=1_comp=128_sum=0.99_ntrain=190000_ntest=10000_nv=17_np=1_NLOS_dftphase=False_down=False.csv};
			
			\addplot[mark options={solid},color=TUMBeamerBlue,line width=1.2pt,mark=o]
			table[x=SNR, y=gmm, col sep=comma]
			{csvdat/2022-10-11_19-42-17_ant_irs=16_ant_bs=8_ant_ue=1_comp=128_sum=0.99_ntrain=190000_ntest=10000_nv=17_np=1_NLOS_dftphase=True_down=False.csv};
			
			\addplot[mark options={solid},color=TUMBeamerBlue,line width=1.2pt,mark=o,dashed]
			table[x=SNR, y=gmm, col sep=comma]
			{csvdat/2022-10-13_05-27-19_ant_irs=16_ant_bs=8_ant_ue=1_comp=128_sum=0.99_ntrain=190000_ntest=10000_nv=17_np=1_NLOS_dftphase=False_down=False.csv};
			
			\addplot[mark options={solid},color=TUMBeamerGreen,line width=1.2pt,mark=square,dotted]
			table[x=snr, y=gmm, col sep=comma]
			{csvdat/cnn_gmm_scen1_1x8BS_1x1UE_4x4IRS_IRS-parallel_Nv=17.csv};
			
			\addplot[mark options={solid},color=TUMBeamerRed,line width=1.2pt,mark=diamond]
			table[x=snr, y=cnn, col sep=comma]
			{csvdat/cnn_gmm_scen1_1x8BS_1x1UE_4x4IRS_IRS-parallel_Nv=17.csv};
			
		\end{axis}
	\end{tikzpicture}
	\caption{$M=8$ ULA BS antennas, $L = 4\times 4 = 16$ \ac{ura} \ac{ris} patches and single-antenna \acp{mt} with $N_v=L+1$.}
	\label{fig:full_scenario1}
\end{figure}
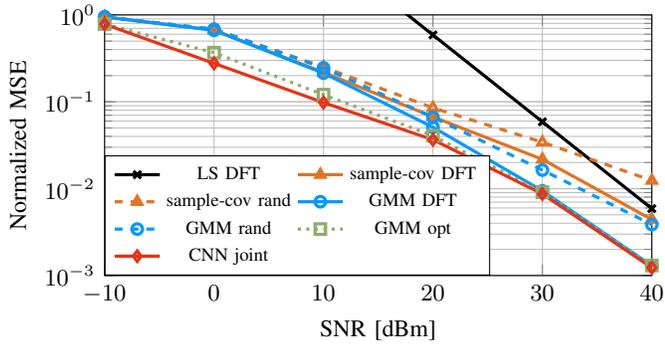

In Fig. \ref{fig:full_scenario1}, we depict results for the case of full illumination, i.e., $N_v=L+1$ with $M=8$ \ac{ula} \ac{bs} antennas and $L=4\times 4$ \ac{ura} \ac{ris} patches. 
In case of the \ac{ls} estimator, the \ac{dft} phase matrix is shown to be optimal, cf. \cite{9747624,9053695}. 
In contrast, a randomly chosen phase matrix might be ill-conditioned in general which results, in combination with the LS estimator, in a normalized \ac{mse} that is larger than one. Therefore, it is not shown in Fig. \ref{fig:full_scenario1}.
When using the \ac{gmm} or the sample-covariance based estimator, the \ac{dft} phase allocations yield a better performance as compared to random allocations. 

Interestingly, it can be observed that the channel estimators with the optimized phase matrix (``CNN joint'' and ``GMM opt'') are able to outperform the \ac{dft} matrix in the low \ac{snr} regime with a vanishing gap in the high \ac{snr}, where the \ac{ls} estimator is reasonable. This means that phase optimization is in fact useful also for the full illumination case for low \ac{snr} values. Furthermore, since the performance is very similar for both the \ac{gmm} and \ac{cnn} estimator it can be concluded that the optimized phase matrix is not only useful for the jointly trained \ac{cnn}, but is generally adapted to the scenario. 

\subsection{Reduced Phase Allocations}\label{sec:numeric_reduced}

	\begin{figure}[t]
	\centering
	\begin{tikzpicture}
		\begin{axis}
			[width=1\columnwidth,
			height=0.57\columnwidth,
			xtick=data, 
			xmin=-10, 
			xmax=50,
			xlabel={SNR [dBm]},
			ymode = log, 
			ymin= 7*1e-3,
			ymax=1e0,
			ylabel= {Normalized MSE}, 
			ylabel shift = 0.0cm,
			grid = both,
			legend columns = 3,
			legend entries={
				\scriptsize LS DFT,
				\scriptsize LS rand,
				\scriptsize sample-cov DFT,
				\scriptsize sample-cov rand,
				\scriptsize GMM DFT,
				\scriptsize GMM rand,
				\scriptsize GMM opt,
				\scriptsize CNN joint,
				\scriptsize GMM hist,
			},
			legend style={at={(0.5,1.0)}, anchor=south},
			]
			
			\addplot[mark options={solid},color=black,line width=1.2pt,mark=x]
			table[x=SNR, y=LS, col sep=comma]
			{csvdat/2022-10-12_09-47-18_ant_irs=16_ant_bs=8_ant_ue=1_comp=128_sum=0.99_ntrain=190000_ntest=10000_nv=8_np=1_NLOS_dftphase=True_down=False.csv};
			
			\addplot[mark options={solid},color=black,line width=1.2pt,dashed,mark=x]
			table[x=SNR, y=LS, col sep=comma]
			{csvdat/2022-10-12_14-31-36_ant_irs=16_ant_bs=8_ant_ue=1_comp=128_sum=0.99_ntrain=190000_ntest=10000_nv=8_np=1_NLOS_dftphase=False_down=False.csv};
			
			\addplot[mark options={solid},color=TUMOrange,line width=1.2pt,mark=triangle]
			table[x=SNR, y=sample_cov, col sep=comma]
			{csvdat/2022-10-12_09-47-18_ant_irs=16_ant_bs=8_ant_ue=1_comp=128_sum=0.99_ntrain=190000_ntest=10000_nv=8_np=1_NLOS_dftphase=True_down=False.csv};
			
			\addplot[mark options={solid},color=TUMOrange,line width=1.2pt,mark=triangle,dashed]
			table[x=SNR, y=sample_cov, col sep=comma]
			{csvdat/2022-10-12_14-31-36_ant_irs=16_ant_bs=8_ant_ue=1_comp=128_sum=0.99_ntrain=190000_ntest=10000_nv=8_np=1_NLOS_dftphase=False_down=False.csv};
			
			\addplot[mark options={solid},color=TUMBeamerBlue,line width=1.2pt,mark=o]
			table[x=SNR, y=gmm, col sep=comma]
			{csvdat/2022-10-12_09-47-18_ant_irs=16_ant_bs=8_ant_ue=1_comp=128_sum=0.99_ntrain=190000_ntest=10000_nv=8_np=1_NLOS_dftphase=True_down=False.csv};
			
			\addplot[mark options={solid},color=TUMBeamerBlue,line width=1.2pt,mark=o,dashed]
			table[x=SNR, y=gmm, col sep=comma]
			{csvdat/2022-10-12_14-31-36_ant_irs=16_ant_bs=8_ant_ue=1_comp=128_sum=0.99_ntrain=190000_ntest=10000_nv=8_np=1_NLOS_dftphase=False_down=False.csv};
			
			\addplot[mark options={solid},color=TUMBeamerGreen,line width=1.2pt,mark=square,dotted]
			table[x=snr, y=gmm, col sep=comma]
			{csvdat/cnn_gmm_scen1_1x8BS_1x1UE_4x4IRS_IRS-parallel_Nv=8.csv};
			
			\addplot[mark options={solid},color=TUMBeamerRed,line width=1.2pt,mark=diamond]
			table[x=snr, y=cnn, col sep=comma]
			{csvdat/cnn_gmm_scen1_1x8BS_1x1UE_4x4IRS_IRS-parallel_Nv=8.csv};
			
			\addplot[mark options={solid},color=mylila,line width=1.2pt,mark=|]
			table[x=snr, y=hist, col sep=comma]
			{csvdat/mse_scen1_hist-snr=40dBm_L=16_ntest=10000_Nv=8.csv};

		\end{axis}
	\end{tikzpicture}
	\caption{$M=8$ ULA BS antennas, $L = 4\times 4 = 16$ \ac{ura} \ac{ris} patches and single-antenna \acp{mt} with $N_v=8$.}
	\label{fig:nv=8_scenario1}
\end{figure}
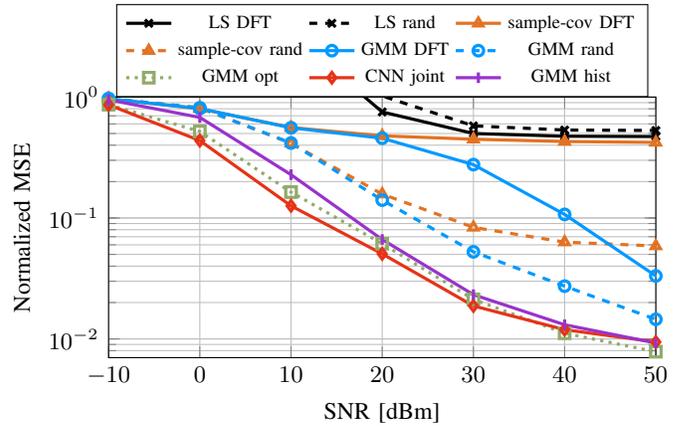

In Fig. \ref{fig:nv=8_scenario1}, we show the same setting as in \Cref{sec:numeric_full} but with a reduced number of $N_v=8$ phase allocations, i.e., less than $50\%$ of the fully illuminated case. First, it can be observed that the \ac{ls} estimator performs poorly due to the under-determined system. 
Since the random phases in combination with the \ac{ls} estimator show no performance gains, we leave out this combination in the following.
Second, the random phase allocations outperform the sub-\ac{dft} allocations when using the \ac{gmm} or sample covariance estimator where the \ac{gmm} estimator performs significantly better than the sample covariance estimator. 
Note that the performance of the sub-\ac{dft} phases is not consistent for an increasing number of phases which can be explained through the insights from \Cref{sec:dft_study}.
Finally, the \ac{cnn} and \ac{gmm} estimator with optimized phase allocations show a similar performance which is better than all baseline methods, including the \ac{gmm} based on the histogram search method from \Cref{sec:dft_study}. This demonstrates the great potential of optimization for reduced phase allocations.

Fig. \ref{fig:phases_scenario1} depicts the same setting as before for a fixed \ac{snr} of 40dBm with a varying number of phase allocations. 
Similarly as before, the methods with optimized phase allocations outperform all baseline algorithms. 
For the \ac{gmm} estimator it is possible to achieve a normalized \ac{mse} of $10^{-2}$ with only $N_v=9$ phase allocations, whereas more than $N_v=12$ phase allocations are needed to achieve the same \ac{mse} when having random or \ac{dft}-based phase allocations.
In the case of full illumination, i.e., $N_v=17$, the \ac{cnn} and \ac{gmm} estimators with optimized or \ac{dft}-based phase allocations show the same performance which is in accordance with Fig. \ref{fig:full_scenario1}.

\begin{figure}[t]
		\centering
		\begin{tikzpicture}
			\begin{axis}
				[width=1\columnwidth,
				height=0.57\columnwidth,
				xtick={8,16,24,32,40,48,56,65},
				xmin=8, 
				xmax=65,
				xlabel={Phases $N_v$},
				ymode = log, 
				ymin= 1e-3,
				ymax=1e0,
				ylabel= {Normalized MSE}, 
				ylabel shift = 0.0cm,
				grid = both,
				legend columns = 3,
				legend entries={
					\scriptsize LS DFT,
					\scriptsize sample-cov DFT,
					\scriptsize sample-cov rand,
					\scriptsize GMM DFT,
					\scriptsize GMM rand,
					\scriptsize GMM opt,
					\scriptsize CNN joint,
				},
				legend style={at={(0.5,1.0)}, anchor=south},
				]
				
				\addplot[mark options={solid},color=black,line width=1.2pt,mark=x]
				table[x=phases, y=LS, col sep=comma]
				{csvdat/2022-10-14_09-24-32_ant_irs=64_ant_bs=16_ant_ue=1_comp=128_sum=0.99_ntrain=190000_ntest=10000_np=1_NLOS_dftphase=True_down=False_phases.csv};
				
				
				\addplot[mark options={solid},color=TUMOrange,line width=1.2pt,mark=triangle]
				table[x=phases, y=sample_cov, col sep=comma]
				{csvdat/2022-10-14_09-24-32_ant_irs=64_ant_bs=16_ant_ue=1_comp=128_sum=0.99_ntrain=190000_ntest=10000_np=1_NLOS_dftphase=True_down=False_phases.csv};
				
				\addplot[mark options={solid},color=TUMOrange,line width=1.2pt,mark=triangle,dashed]
				table[x=phases, y=sample_cov, col sep=comma]
				{csvdat/2022-10-19_02-38-40_ant_irs=64_ant_bs=16_ant_ue=1_comp=128_sum=0.99_ntrain=190000_ntest=10000_np=1_NLOS_dftphase=False_down=False_phases.csv};
				
				\addplot[mark options={solid},color=TUMBeamerBlue,line width=1.2pt,mark=o]
				table[x=phases, y=gmm, col sep=comma]
				{csvdat/2022-10-14_09-24-32_ant_irs=64_ant_bs=16_ant_ue=1_comp=128_sum=0.99_ntrain=190000_ntest=10000_np=1_NLOS_dftphase=True_down=False_phases_gmm.csv};
				
				\addplot[mark options={solid},color=TUMBeamerBlue,line width=1.2pt,mark=o,dashed]
				table[x=phases, y=gmm, col sep=comma]
				{csvdat/2022-10-19_02-38-40_ant_irs=64_ant_bs=16_ant_ue=1_comp=128_sum=0.99_ntrain=190000_ntest=10000_np=1_NLOS_dftphase=False_down=False_phases.csv};
				
				\addplot[mark options={solid},color=TUMBeamerGreen,line width=1.2pt,mark=square,dotted]
				table[x=nv, y=gmm, col sep=comma]
				{csvdat/cnn_gmm_scen3_1x16BS_1x1UE_8x8IRS_IRS-parallel_snr=40dBm.csv};
				
				\addplot[mark options={solid},color=TUMBeamerRed,line width=1.2pt,mark=diamond]
				table[x=nv, y=cnn, col sep=comma]
				{csvdat/cnn_gmm_scen3_1x16BS_1x1UE_8x8IRS_IRS-parallel_snr=40dBm.csv};
				
			\end{axis}
		\end{tikzpicture}
		\caption{$M=16$ ULA BS antennas, $L = 8\times 8 = 64$ URA \ac{ris} patches and single-antenna \acp{mt} with $\text{SNR}=40$dBm.}
		\label{fig:phases_scenario3}
	\end{figure}
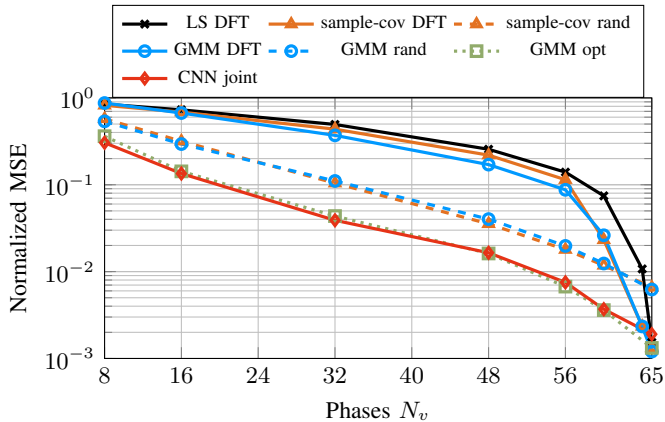

Finally, in Fig. \ref{fig:phases_scenario3}, we show results for a larger system setup with $M=16$ \ac{ula} \ac{bs} antennas and $L=8\times 8$ \ac{ura} \ac{ris} patches for a fixed \ac{snr} of 40dBm. It can be observed that especially the gap to the \ac{dft}-based phase allocations increases drastically which perform poorly for this larger system setup. 
Once again, the optimized phase allocations allow for drastic performance gains, which is equivalent to requiring less phase allocations to achieve the same estimation quality.
In conclusion, the optimization of the phase allocations has increasing potential for larger systems which is in compliance with the trend to massive \ac{mimo} systems.

\begin{figure}[t]
	\centering
	\begin{tikzpicture}
		\begin{axis}
			[width=1\columnwidth,
			height=0.57\columnwidth,
			xtick={2,5,8,11,14,17}, 
			xmin=2, 
			xmax=17,
			xlabel={Phases $N_v$},
			ymode = log, 
			ymin= 1e-3,
			ymax=1e0,
			ylabel= {Normalized MSE}, 
			ylabel shift = 0.0cm,
			grid = both,
			legend columns = 3,
			legend entries={
				\scriptsize LS DFT,
				\scriptsize sample-cov DFT,
				\scriptsize sample-cov rand,
				\scriptsize GMM DFT,
				\scriptsize GMM rand,
				\scriptsize GMM opt,
				\scriptsize CNN joint,
			},
			legend style={at={(0.5,1.0)}, anchor=south},
			]
			
			\addplot[mark options={solid},color=black,line width=1.2pt,mark=x]
			table[x=phases, y=LS, col sep=comma]
			{csvdat/2022-10-12_10-06-46_ant_irs=16_ant_bs=8_ant_ue=1_comp=128_sum=0.99_ntrain=190000_ntest=10000_np=1_NLOS_dftphase=True_down=False_phases.csv};
			
			
			\addplot[mark options={solid},color=TUMOrange,line width=1.2pt,mark=triangle]
			table[x=phases, y=sample_cov, col sep=comma]
			{csvdat/2022-10-12_10-06-46_ant_irs=16_ant_bs=8_ant_ue=1_comp=128_sum=0.99_ntrain=190000_ntest=10000_np=1_NLOS_dftphase=True_down=False_phases.csv};
			
			\addplot[mark options={solid},color=TUMOrange,line width=1.2pt,mark=triangle,dashed]
			table[x=phases, y=sample_cov, col sep=comma]
			{csvdat/2022-10-12_10-06-46_ant_irs=16_ant_bs=8_ant_ue=1_comp=128_sum=0.99_ntrain=190000_ntest=10000_np=1_NLOS_dftphase=False_down=False_phases.csv};
			
			\addplot[mark options={solid},color=TUMBeamerBlue,line width=1.2pt,mark=o]
			table[x=phases, y=gmm, col sep=comma]
			{csvdat/2022-10-12_10-06-46_ant_irs=16_ant_bs=8_ant_ue=1_comp=128_sum=0.99_ntrain=190000_ntest=10000_np=1_NLOS_dftphase=True_down=False_phases.csv};
			
			\addplot[mark options={solid},color=TUMBeamerBlue,line width=1.2pt,mark=o,dashed]
			table[x=phases, y=gmm, col sep=comma]
			{csvdat/2022-10-12_10-06-46_ant_irs=16_ant_bs=8_ant_ue=1_comp=128_sum=0.99_ntrain=190000_ntest=10000_np=1_NLOS_dftphase=False_down=False_phases.csv};
			
			\addplot[mark options={solid},color=TUMBeamerGreen,line width=1.2pt,mark=square,dotted]
			table[x=nv, y=gmm, col sep=comma]
			{csvdat/cnn_gmm_scen1_1x8BS_1x1UE_4x4IRS_IRS-parallel_snr=40dBm.csv};
			
			\addplot[mark options={solid},color=TUMBeamerRed,line width=1.2pt,mark=diamond]
			table[x=nv, y=cnn, col sep=comma]
			{csvdat/cnn_gmm_scen1_1x8BS_1x1UE_4x4IRS_IRS-parallel_snr=40dBm.csv};

		\end{axis}
	\end{tikzpicture}
	\caption{$M=8$ ULA BS antennas, $L = 4\times 4 = 16$ \ac{ura} \ac{ris} patches and single-antenna \acp{mt} with $\text{SNR}=40$dBm.}
	\label{fig:phases_scenario1}
\end{figure}

\section{Conclusion}
In this work, we investigated the potential of optimizing the reduced phase allocation matrix for channel estimation in \ac{ris}-aided systems which tackles the problem of unaffordable large pilot overhead for full illumination. With a study based on a selection of \ac{dft} columns, we found that the system setup drastically influences the choice of the optimal phase allocations. We proposed a \ac{nn} which jointly learns a phase allocation matrix together with a channel estimator. 
The proposed approach outperforms the baseline approaches over the whole \ac{snr} range by a large margin. In addition, when using the optimized phase allocation matrix for a different state-of-the-art channel estimator, its performance is significantly increased. 
This leads to the conclusion that the optimized phase allocation matrix is able to leverage the inherent structure of the \ac{bs}' environment to performance gains. 


	
	\bibliographystyle{IEEEtran}
	\bibliography{IEEEabrv,bibliography}

\end{document}

%% file: befe.tex
\usepackage{mathtools}	
\usepackage{amssymb}
\usepackage{amsthm}

\usepackage[utf8]{inputenc}

\usepackage{ifthen}

\usepackage{microtype}

\usepackage{caption}

\usepackage{bm}

\usepackage[draft,nomargin,marginclue,inline]{fixme}
\makeatletter
\renewcommand*\FXLayoutMarginClue[3]{%
	\marginpar[%
	\raggedleft\@fxuseface{margin}\textcolor{red}{\ignorespaces $ \Rightarrow $}]{%
		\raggedright\@fxuseface{margin}\textcolor{red}{\ignorespaces $ \Leftarrow $}}}
\makeatother	

\usepackage{tumcolor}

\usepackage{pgfplotstable}	

\pgfplotsset{
	discard if/.style 2 args={
		x filter/.append code={
			\edef\tempa{\thisrow{#1}}
			\edef\tempb{#2}
			\ifx\tempa\tempb
			
			\fi
		}
	},
	discard if not/.style 2 args={
		x filter/.append code={
			\edef\tempa{\thisrow{#1}}
			\edef\tempb{#2}
			\ifx\tempa\tempb
			\else
			
			\fi
		}
	}
}

\usepackage{cite}
\usepackage{hyperref}
\hypersetup{
	colorlinks = true,
	linkbordercolor = {white},
	citecolor = {black},
	linkcolor = {black},
	urlcolor = {black}
}
\makeatletter
\renewcommand*{\eqref}[1]{%
	\hyperref[{#1}]{\textup{\tagform@{\ref*{#1}}}}%
}
\makeatother

\usepackage[acronym,shortcuts]{glossaries}
\newacronym{blmmse}{BLMMSE}{Bussgang LMMSE}
\newacronym{cnn}{CNN}{convolutional neural network}
\newacronym{dft}{DFT}{discrete Fourier transform}
\newacronym{em}{EM}{expectation-maximization}
\newacronym{emiht}{EM-IHT}{EM algorithm with IHT}
\newacronym{iht}{IHT}{iterative hard thresholding}
\newacronym{lmmse}{LMMSE}{linear minimum mean square error}
\newacronym{mimo}{MIMO}{multiple-input multiple-output}
\newacronym{mmse}{MMSE}{minimum mean square error}
\newacronym{mse}{MSE}{mean square error}
\newacronym{relu}{ReLU}{rectified linear unit}
\newacronym{snr}{SNR}{signal-to-noise ratio}
\newacronym{adc}{ADC}{analog-to-digital converter}
\newacronym{nn}{NN}{neural network}
\newacronym{re}{RE}{resource element}

\newcommand{\op}[1]{{\operatorname{#1}}}
\newcommand{\uproman}[1]{\uppercase\expandafter{\romannumeral#1}}

\newcommand{\vect}{\operatorname{vec}}
\newcommand{\diag}{\operatorname{diag}}

\newcommand{\eye}{\bm{\op{I}}}
\newcommand{\B}[1]{\bm{#1}}
\newcommand{\inv}{^{-1}}

\newcommand{\h}{^{\op H}}
\newcommand{\T}{^{\op T}}

\usepackage{cleveref}

\usepackage{enumitem}

\newcommand{\plotwidth}{0.95\columnwidth}

\newcommand{\lineWidth}{1.0pt}
\tikzset{algorithm1/.style={mark options={solid},color=TUMBeamerBlue, line width=\lineWidth, mark=square, dashed}}

\newcommand*{\C}{\mathbb{C}}

\newcommand*{\R}{\mathbb{R}}

\newlength{\leftstackrelawd}
\newlength{\leftstackrelbwd}
\def\leftstackrel#1#2{\settowidth{\leftstackrelawd}%
	{${{}^{#1}}$}\settowidth{\leftstackrelbwd}{$#2$}%
	\addtolength{\leftstackrelawd}{-\leftstackrelbwd}%
	\leavevmode\ifthenelse{\lengthtest{\leftstackrelawd>0pt}}%
	{\kern-.5\leftstackrelawd}{}\mathrel{\mathop{#2}\limits^{#1}}}